\documentclass[prb,twocolumn,aps,showpacs,fixfloats]{revtex4}
\usepackage{graphicx}
\usepackage{bm}
\usepackage{amsmath,amssymb}
\usepackage{subfigure}
\usepackage{float}
\usepackage{latexsym}
\usepackage{epstopdf}
\usepackage{color}
\usepackage{enumerate}
\usepackage{pdfpages}

\newcommand{\s}{\scriptscriptstyle}

\newcommand{\ket}[1] { \left|{#1}\right> }
\newcommand{\bra}[1] { \left<{#1}\right| }

\renewcommand{\u}{\uparrow}

\begin{document}

\title {Spin transport with traps: dramatic narrowing of the Hanle curve }
 \affiliation{ Department of Physics and
Astronomy, University of Utah, Salt Lake City, UT 84112, USA}

\begin{abstract}
We study theoretically the spin transport in a device in which
the active layer
%, sandwiched between the magnetized electrodes,
is an organic film with numerous deep in-gap levels
serving as traps. A carrier,
diffusing between magnetized injector and detector, spends a considerable
portion of time on the traps. This new feature of transport does not affect the giant magnetoresistance, which is sensitive only to the mutual orientation of
magnetizations of the injector and detector. By contrast,
the presence of traps
strongly affects
%modifies
%dramatically
the sensitivity of
the spin transport
%nonlocal
%resistance to
%the
to external magnetic field
perpendicular to the magnetizations of the electrodes
%magnetic field
(the Hanle effect). Namely,
the Hanle curve narrows
%drastically
dramatically. The origin of such a narrowing is that the
spin precession takes place during the {\em entire} time of the carrier motion between the electrodes, while the spin relaxation takes place only  during diffusive motion between the subsequent traps.  If the resulting width of the Hanle curve is smaller than
the measurement resolution,
%unavoidable uncertainty in the actual value of magnetic field, the
observation of the Hanle peak becomes impossible.
\end{abstract}

\maketitle

\section{Introduction}
\begin{figure}
\includegraphics[width=90mm]{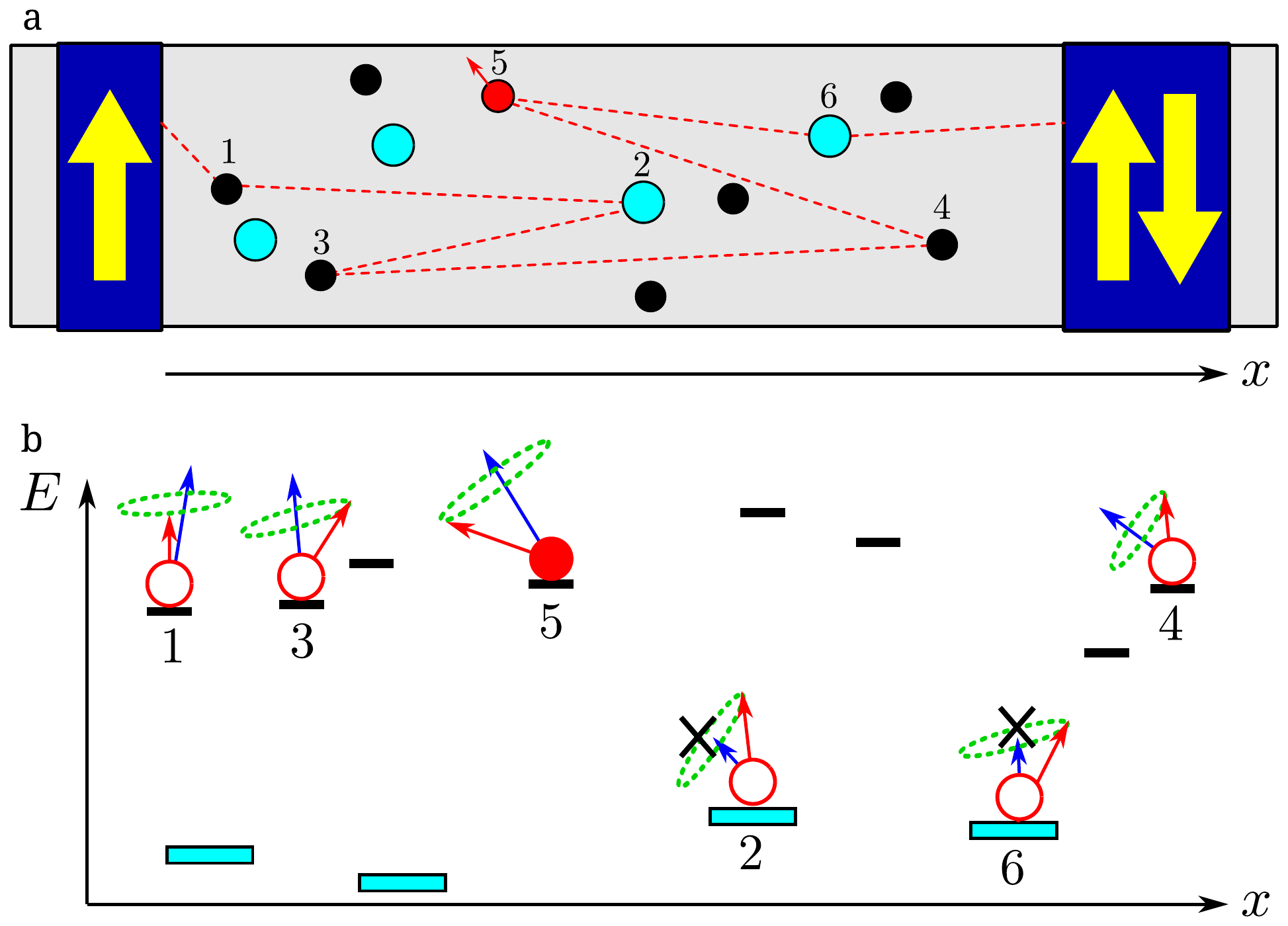}
\caption{(Color Online). Carrier transport between the magnetized electrodes in the presence of deep traps is illustrated
schematically in coordinate (a) and energy (b) spaces.  An injected spin-up carrier
relaxes the spin while visiting the sites numbered as $1$, $3$, $4$, and $5$, but {\em preserves} the spin while visiting the traps numbered as $2$ and $6$.}   \label{fig1}
\end{figure}
Observation of the giant magnetoresistance (GMR) effect in organic
devices\cite{OrganicValve1,OrganicValve1'} was later
reproduced by many groups
on various
organic active layers and with various ferromagnetic electrodes, see e.g. Refs. \onlinecite{Valve1,Valve2,Valve3,Valve4,Valve5,Valve6}.
Along with demonstration of GMR, the value of
spin diffusion length in organic film, $\lambda_s=40$nm,   was inferred in Ref. \onlinecite{OrganicValve1'} from the thickness dependence of the effect.
This value is by a factor of  $\sim 20$ smaller than  $\lambda_s$
in a number of conventional semiconductors, see e.g. Refs. \onlinecite{Conventional1,Conventional2,Conventional3}.
These and many other papers where the GMR effect is reported, also report
the observation of the Hanle effect.
%response.
It is the latter observation which constitutes an  unambiguous proof that the actual spin transport between the electrodes
takes place. The Hanle effect manifests itself as a drop of the resistance of the
structure with channel length,
$L\sim \lambda_s$,
%on the order of $\lambda_s$
as the external
field normal to the magnetizations of the electrodes is applied. This drop is the
 result of the Larmor precession of spins of  the injected carriers.
%of the actual spin transport
%between the electrodes. This proof is
%
%
%that the actual spin transport
%takes place
%In Refs. \onlinecite{Conventional1,Conventional2,Conventional3} and many
%other experimental papers where the GMR effect is reported, the
%%these and many other papers
%the fact that
%%observed
%%GMR
%it is due to
%actual spin transport was
%unambiguously confirmed by
%the detection of the Hanle
%response.
%Specifically, the
%resistance of the structures with channel length,
%$L$, on the order of $\lambda_s$ was shown to drop with external
% field normal to the magnetization of electrodes. This drop is the
% result of the Larmor precession of spins of  the injected carriers.

Despite indirect indications\cite{InjectionYes1,InjectionYes2,InjectionYes3}
of a finite spin polarization in the active layer,
the Hanle effect in organic devices is either completely missing\cite{NoInjection1,NoInjection2,NoInjection3}
or shows up as a
weak signature\cite{HanleOrganicJapanese1,HanleOrganicJapanese2}.
In experimental papers Refs. \onlinecite{NoInjection1,NoInjection2,NoInjection3}, %\onlinecite{NoInjection2}
the puzzling absence of the Hanle effect was ascribed to a strong inhomogeneity
of either organic layer itself\cite{NoInjection1,NoInjection2} or of the electrodes\cite{NoInjection3}.
In a theoretical paper, Ref. \onlinecite{Yu}, the explanation of the ``missing" Hanle effect
%As possible explanation of the absence of the Hanle effect
%it was suggested that the transport between the electrodes is due
%to tunneling between the electrodes via the pinholes in organic %layer\cite{NoInjection1, NoInjection2}.  Alternative explanation\cite{NoInjection2} %or tilting out of plane of the
%magnetization of the electrodes.
%,that it is completely smeared out as a result of inhomogeneity of magnetization of %electrodes\cite{NoInjection3}.
%Yet another explanation\cite{Yu}
dwells upon a presumed
specific property
%feature
of organic materials, namely, strong exchange coupling
between carriers which  leads  to anomalously short spin diffusion time, $\tau_s$.
According to Ref. \onlinecite{Yu}, short $\tau_s$ requires very strong magnetic fields to reveal the spin precession. In other words, the explanation %\cite{Yu}
of the absence of the Hanle effect is that the Hanle curve is {\em too broad}.

In the present paper we exploit a different intrinsic property of organic
semiconductors which distinguishes them from the conventional crystalline
semiconductors. This property is the presence of
%numerous
deep traps, see Fig. \ref {fig1}, which
a carrier visits on the way between the injector and detector.
Our only assumption about these traps is that, while sitting on a trap,
a carrier is not subject to spin relaxation. From this assumption we readily
derive that, while the GMR response is unaffected by  traps,
the Hanle effect is affected dramatically. Namely, as a result of visiting the traps, the Hanle curve {\em narrows}. This scenario, although opposite to
Ref. \onlinecite{Yu}, also inhibits the observability of the Hanle effect.
The effect will not be detectable if the  width of the Hanle curve
is smaller than the measurement resolution.

\begin{figure}
\includegraphics[width=77mm]{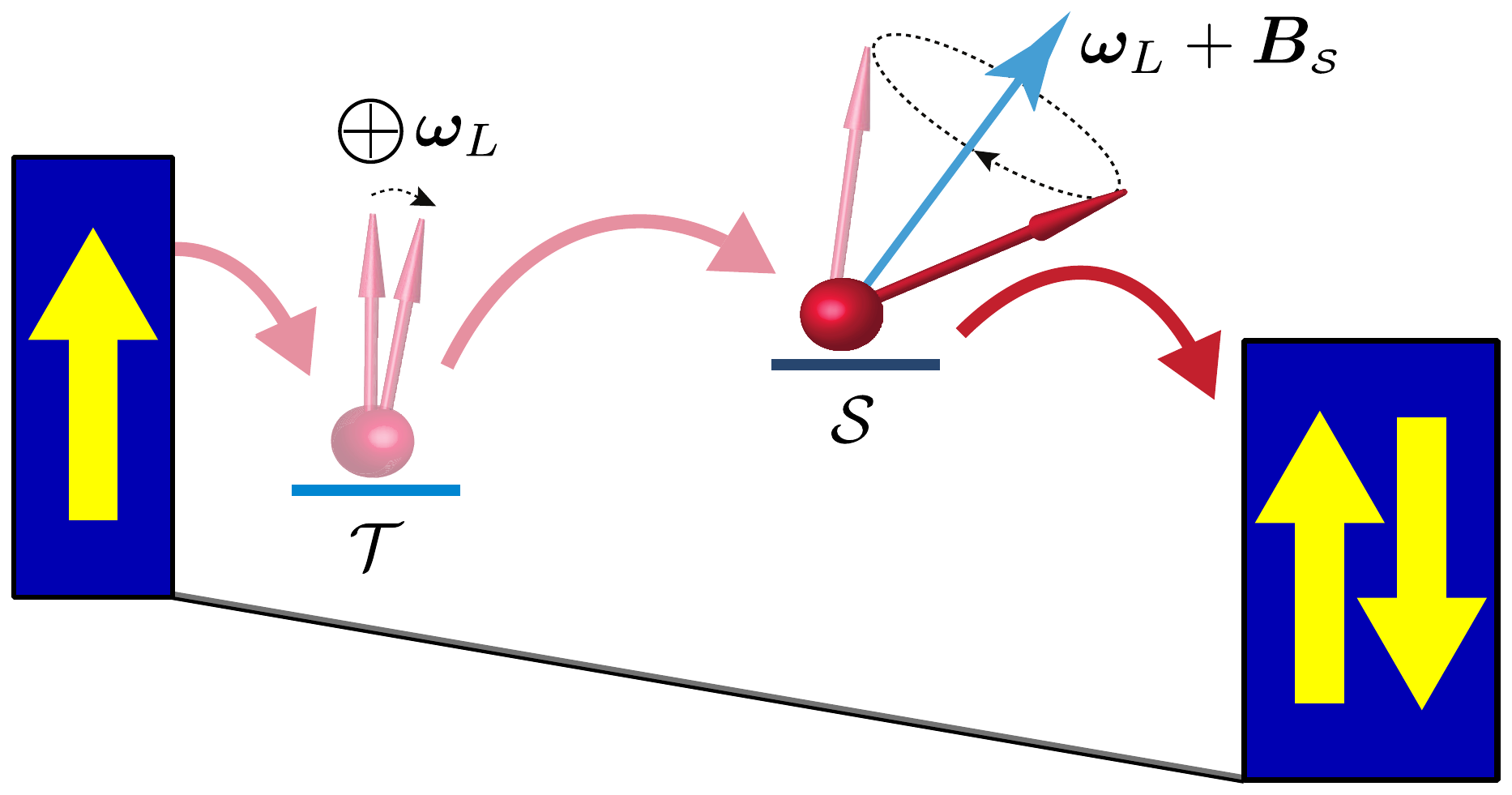}
\caption{(Color Online). A cartoon model of a two-step transport between the magnetized electrodes. The intermediate states are: a trap, ${\cal T}$, with a long waiting time and no local field, and a site ${\cal S}$, which hosts the field ${\bm B}_{\s {\cal S}}$. The trap dominates the Hanle response in external field, ${\bm \omega}_L$, while the site fully controls the value of GMR.
}
\label{fig2}
\end{figure}

%\section{A toy model}
\noindent{\em A toy model}. To illustrate our message, consider
a toy
%the simplest
model of GMR in organics\cite{multistep1,multistep2} illustrated in Fig. \ref{fig2}.
The current between the electrodes
%with polarizations ${\cal P}_1$ and
%${\cal P}_2$
is due to a sequential hopping via only two intermediate states, ${\cal T}$ and ${\cal S}$.
Denote with ${\bm B}_{\s {\cal S}}$ and ${\bm B}_{\s {\cal T}}$ the on-site fields in which the carrier spin precesses while waiting for the hop. The advantage of this model is that the Hanle signal,
defined as\cite{Silsbee1988,vanWeesPioneering,Crowell2007}
\begin{equation}
\label{general}
R_H \propto \int_0^{\infty}dt S_z(t),
\end{equation}
can be calculated explicitly. With two steps, the expression for $R_H$ can be cast in the form
\begin{align}
\label{DoubleIntegral}
 R_H&=C_2\int\limits_0^\infty dt_{\s {\cal S}} f_{\s {\cal S}}(t_{\s {\cal S}}) \int\limits_0^\infty dt_{\s {\cal T}} f_{\s {\cal T}}(t_{\s {\cal T}}) \nonumber \\
& \times \left\{ \left| \bra{\u} \widehat{U}({\bm B}_{\s {\cal S}}, t_{\s {\cal S}}) \widehat{U}({\bm B}_{\s {\cal T}}, t_{\s {\cal T}}) \ket{\u} \right|^2
-\frac{1}{2}
\right\},
\end{align}
 where $f_{\s {\cal S}}(t_{\s {\cal S}})$ and  $f_{\s {\cal T}}(t_{\s {\cal T}})$ are the distribution functions of the waiting times, $t_{\s {\cal T}}$ and  $t_{\s {\cal S}}$, and $\widehat{U}$ is the evolution operator,
$\widehat{U}({\bm B}, t) = \exp\left[-it \left({\bm B}{\bm S}\right)\right]$,
in a magnetic field, ${\bm B}$. Straightforward  evaluation of the double integral in Eq. (\ref{DoubleIntegral})
yields\cite{multistep2}
\begin{multline}
\label{DoubleIntegral1}
R_H= \frac{C_2}{2}\left\{ \left( \frac{1 + B_{{\s {\cal T}}z}^2 \tau_{\s {\cal T}}^2 }{1
 + B_{\s {\cal T}}^2 \tau_{\s {\cal T}}^2 }
\right)
 \left( \frac{1 + B_{{\s {\cal S}}z}^2 \tau_{\s {\cal S}}^2 }{1
 + B_{\s {\cal S}}^2 \tau_{\s {\cal S}}^2 }
\right) \right. \\
\left.
- \text{Re}\left[ \frac{
B_{ {\s {\cal T}} +} B_{ {\s {\cal S}} -} \tau_{\s {\cal T}}
 \tau_{\s {\cal S}}
(1 + i B_{ {\s {\cal T}} z} \tau_{\s {\cal T}})
(1 - i B_{ {\s {\cal S}} z} \tau_{\s {\cal S}})
}{(1
 + B_{\s {\cal T}}^2 \tau_{\s {\cal T}}^2 ) (1
 + B_{\s {\cal S}}^2 \tau_{\s {\cal S}}^2)}
\right] \right\},
\end{multline}
%Here $\tau_{\s {\cal S}}$ and  $\tau_{\s {\cal T}}$,
where ${B}_{\pm}= B_x \pm i B_y$, and $\tau_{\s {\cal S}}$,
and $\tau_{\s {\cal T}}$ are the average waiting times.

We now specify ${\cal S}$ as a site and ${\cal T}$ as a trap.
Namely, the site hosts a random magnetic field, and mimics the
spin relaxation in the course of  charge transport.
%On the other hand
The specifics of the trap, ${\cal T}$, %represents a trap, so that
is that the spin is {\em not} rotated
when a charge is on ${\cal T}$, and also the waiting time,  $\tau_{\s {\cal T}}$,
is much longer than $\tau_{\s {\cal S}}$. In a weak external field, ${\bm \omega}_L$, directed along the $x$-axis
we have ${\bm B}_{\s {\cal S}}\rightarrow {\bm B}_{\s {\cal S}}+{\bm \omega}_L$ and
${\bm B}_{\s {\cal T}}={\bm \omega}_L$.

Upon inspection of Eq. (\ref {DoubleIntegral1}), one can conclude that  with only two conditions: ({\em i}) ${\omega}_L \ll B_{\s {\cal S}}$ and ({\em ii}) $\tau_{\s {\cal T}}\gg\tau_{\s {\cal S}}$ satisfied, the expression for $R_H$, averaged over the in-plane orientations of ${\bm B}_{\s {\cal S}}$,
simplifies to
\begin{equation}
\label{simplified}
R_H(\omega_L) = \frac{C_2}{2} \frac{1 + B_{{\s {\cal S}} z}^2 \tau_{\s {\cal S}}^2 }{
(1+ \omega_L^2 \tau_{\s {\cal T}}^2)(1+ B_{\s {\cal S}}^2 \tau_{\s {\cal S}}^2) }.
\end{equation}
We see that, as a function of external field, the Hanle signal is a Lorentizan with a width
determined {\em exclusively} by the time spent on the trap, $\tau_{\s {\cal T}}$.
Note also, that, in the absence of external field, Eq. (\ref{DoubleIntegral1}) yields
\begin{equation}
\label{ZeroField}
R_H(0)= \frac{C_2}{2} \left( \frac{1 + B_{{\s {\cal S}}z}^2 \tau_{\s {\cal S}}^2 }{1
 + B_{\s {\cal S}}^2 \tau_{\s {\cal S}}^2 }
\right),
\end{equation}
i.e. the  value which does not ``know" about the trap. On the other hand, it is this value
that is responsible for the GMR. This follows from the realization that GMR
%is because the
%magnitude of GMR
%%GMR
is determined by the probability
to {\em preserve} spin during the travel between the magnetized electrodes.
The structure of Eq.~(\ref{DoubleIntegral}) suggests that this preservation
probability has the same form only without the prefactor $C_2$ and without $1/2$
in the integrand.
%This probability is {\em also} given by  Eq.~(\ref{DoubleIntegral})  (without the prefactor $C_2$) at $\omega_L=0$.
Certainly,
such a direct relation is due to the simplicity of our toy model.

We have illustrated how the presence of a trap leads to a ``decoupling" of the Hanle effect
from the GMR. In the next section we calculate the Hanle profile for a more realistic setup, when a carrier diffuses between the subsequent traps.

%\section{Hanle lineshape in the presence of deep traps}
\noindent{\em Hanle lineshape in the presence of deep traps}.
The central notion behind the Hanle effect is that the contribution to nonlocal resistance from a carrier injected at time $t=0$ with spin
directed the along the $x$-axis precesses with time as $\cos\omega_Lt$.
%, where $\omega_L$ is the Larmour frequency.
The standard Hanle profile emerges upon
%averaging of this
summation of all these contributions
\begin{equation}
\label{ClassicHanle}
R(\omega_L) = C \int_0^\infty dt \cos\left( \omega_L t \right) e^{-t/\tau_s} P_L(t).
\end{equation}
Here the weighting factor,
\begin{equation}
P_L(t) = \frac{1}{(4 \pi D t)^{1/2}} \exp\left[ -\frac{L^2 }{4 D t}\right],
\end{equation} takes into
account that the electron travels to the detector at $x=L$ diffusively, while the factor $\exp\left(-t/\tau_s\right)$
describes the spin memory loss with a constant rate $\tau_s^{-1}$.
Incorporation of traps requires the following modification of Eq. (\ref{ClassicHanle}). The
spin precession takes place {\em both} during the time, $t$, spent in course of diffusion, and the time $t_{\s tr}$
spent  while sitting on the traps. In other words, $\cos \omega_Lt$ should be modified as follows
\begin{equation}
\label{traps}
\cos\omega_Lt \rightarrow \Big< \left< \, \cos \omega_L(t + t_{\s tr})\right>_{t_{\s tr}}\Big>_{\{x_i\}}.
\end{equation}
Here the first averaging is performed over the waiting times spent on traps for fixed coordinates  of the traps, while the subscript
$\{x_i\}$ stands for the positional averaging or, more precisely, for averaging over the positions of the traps that
a carrier encounters along its way from injector to detector.
Obviously, the order in which the averaging in Eq. (\ref{traps}) is performed is important.  This is because the
first averaging presumes that the number, $n$,  of encountered traps is fixed. For this fixed $n$ the
averaging over $t_{\s tr}$ reduces to the $n$-fold integral
\begin{equation}
\label{withdeltafunction}
  \prod\limits_{j=1}^n \int\limits_0^\infty dt_j f_j(t_j) \text{Re} \left\{ \exp\left[i\omega_L\left(t+ \sum\limits_{j=1}^n t_j\right) \right] \right\},
\end{equation}
where $f_j(t_j)$ is the distribution function of the random times, $t_j$, spent on $j$-th trap. This time encapsulates the waiting for trapping and waiting for the release. Since the second %contribution is much bigger,
process is much slower,
the distribution is Poissonian
\begin{equation}
\label{Poissonian}
f_j(t_j) = \frac{1}{\tau_j} \exp{\Bigl(-\frac{t_j}{\tau_j}\Bigr)},
\end{equation}
where $\tau_j$ is the characteristic waiting time for release from the $j$-th trap.
With the help of this distribution we readily obtain
%{\color{red}
%Using the representation of the $\delta$-function
%$\delta(z)=\frac{1}{2\pi}\int d\kappa \exp(i\kappa z)$, the  $(n+1)$-fold integral Eq. (\ref{withdeltafunction}) can be reduced to a
%two-fold integral. }
%We will adopt another simplification based on the following qualitative argument. The dramatic modification of the Hanle profile by the traps occurs when the net time spent on the traps is longer than the time spent in motion. Thus, we can
%neglect   $t$ in the argument of cosine compared to $t_{\s tr}$. This assumption, which we will check later, allows
%to simplify $\langle \cos(\omega_L (t +t_{\s tr}) \rangle_{t_{\s tr}}$ to
\begin{equation}
\label{product}
\langle \cos(\omega_L (t +t_{\s tr}) \rangle_{t_{\s tr}} \approx
\text{Re} \left\{ \exp(i \omega_L t) \prod_{j=1}^n \frac{1}{1-i \omega_L \tau_j}
\right\}.
\end{equation}
As a next crucial step, we take into account that the trap levels are distributed within a wide interval, so that the  waiting times, $\tau_j$,
are widely dispersed. Conventionally, see e.g. Refs.
\onlinecite{WaitingTimesClassic}, \onlinecite{FlatteWaitingTimes1},
 their distribution
is modeled by a function $p(\tau_j)$ which falls off as a power law $\sim \tau_j^{-\alpha}$ at large $\tau_j$ and is flat for small $\tau_j$. Such distributions are called heavy-tailed in the literature and are characterized by a divergent mean.
For concreteness, we will carry out the calculations for the Lorentzian distribution
\begin{equation}
p(\tau_j) = \frac{2}{\pi} \frac{\tau_0}{\tau_j^2 + \tau_0^2},
\end{equation}
corresponding to $\alpha=2$ and cutoff $\tau_0$.

Upon averaging with $p(\tau_j)$, each factor in the product in the integrand of Eq. (\ref{product})
assumes the form
\begin{equation}
\left<\frac{1}{1 - i \omega_L \tau_{j}} \right>_{\tau_j}
%= \frac{1}{1- \kappa^2 \tau_0^2} \left( 1 +\kappa \tau_0 + i \frac{2}{\pi} \kappa  \tau_0 \ln \kappa \tau_0\right).
= \frac{1}{1- \omega_L^2 \tau_0^2} \left( 1 - \omega_L \tau_0 + i \frac{2}{\pi} \omega_L  \tau_0 \ln|\omega_L \tau_0|\right),
\end{equation}
%Note that, upon substituting $\Bigl[\left<\frac{1}{1 + i \kappa \tau_{j}} \right>_{\tau_j}\Bigr]^n$ into the integrand of
%Eq. (\ref{product}), both integrations can be easily performed. This is because the integral over $t_{\s tr}$ reduces
%to the $\delta$-function and sets $k=-\omega_L$. 
which leads to the closed analytical expression for $\langle \cos(\omega_L (t +t_{\s tr}) \rangle_{t_{\s tr}}$
\begin{widetext}
\begin{equation}
\label{wide}
\left< \langle \cos(\omega_L (t +t_{\s tr}) \rangle_{t_{\s tr}} \right>_{\{x_i\}} =
\left<
    \frac{1}{(1 + \omega_L \tau_0)^n}
    \left( 1 + \frac{4 \omega_L^2 \tau_0^2 \ln^2 \omega_L \tau_0}{\pi^2(1-\omega_L \tau_0)^2} \right)^{n/2}
    \cos\left[ \omega_L t + n \tan^{-1} \frac{2}{\pi} \frac{\omega_L \tau_0 \ln \omega_L \tau_0}{1 - \omega_L \tau_0}  \right]
\right>_{\{x_i\}}.
\end{equation}
\end{widetext}
Averaging over $x_i$ in Eq. (\ref{wide}) must be understood as follows. For a given set of the coordinates of
traps different diffusion trajectories can visit different number of traps. It is a delicate issue that, depending on the diffusion time, $t$, in the argument of Eq. (\ref{ClassicHanle}), the number,  $n$, of the {\em visited} traps
 is different. This issue is intimately related to the specifics of a random walk which is accompanied by multiple returns to each site visited previously.
%By contrast,
Clearly, for unidirectional drift, the number of visited traps is $n={\cal N}L$, where ${\cal N}$ is the density of traps, regardless of the travel time. For a  diffusion motion, the dependence of $n$ on ${\cal N}$ is superlinear. Indeed,
with traps homogeneously distributed along the carrier path,
%What we can  assert with confidence, is that, for , the number,
$n$ is proportional to the time $t$ during which the carrier diffuses.  
%the net time, $t$.
%provided that the traps are homogeneously distributed.
In other words
 \begin{equation}
 \label{InOtherWords}
 n(t)=\frac{t}{\tau^{\ast}},~~~~~\tau^{\ast}=\frac{1}{D{\cal N}^2}.
 \end{equation}
 The physical meaning of $\tau^{\ast}$ is the diffusion time between the neighboring traps.
 The true numerical factor in Eq.~(\ref{InOtherWords})  cannot be specified
 by such a simple reasoning.

\begin{figure}
\includegraphics[width=77mm]{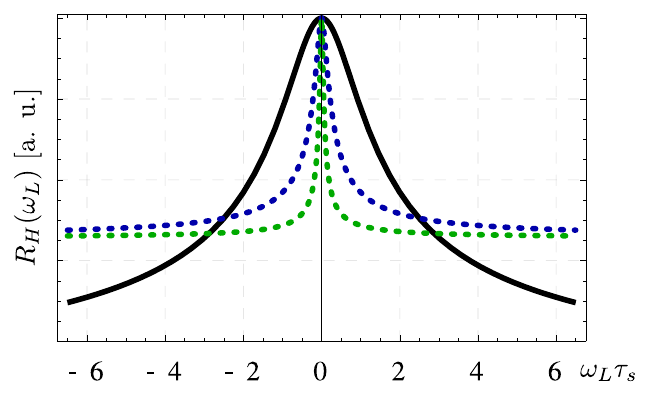}
\caption{(Color Online). Evolution of the Hanle response, $R_H(\omega_L)$, with the density
of traps, ${\cal N}$, measured in the units $(D\tau_0)^{-1/2}$, is plotted from Eqs. (\ref{InOtherWords}),
(\ref{asymptotic}), and (\ref{nontrivial}). Three curves correspond to  ${\cal N} = 0$ (black), ${\cal N} =(2/D\tau_0)^{1/2}$ (blue), and ${\cal N} = 2(2/D\tau_0)^{1/2}$ (green).
 Characteristic density, $(D\tau_0)^{-1/2}$, correspons to one trap per diffusion displacement during the trapping time.  }
\label{fig3}
\end{figure}

 The remaining task is to substitute Eq. (\ref{wide}) with $n$ given by Eq. (\ref{InOtherWords}) into Eq.
 (\ref{ClassicHanle}) and to perform integration over time. As we will see later, the characteristic width
 of the Hanle curve in the presence of traps is much smaller than a typical trapping time. This allows us to
 expand Eq. (\ref{wide}) with respect to a small parameter $\omega_L\tau_0$. The resulting expression for $R_H$
 %reads
 takes a simple form
\begin{widetext}
\begin{equation}
\label{resulting}
R_H(\omega_L) = \int_0^\infty dt \; \frac{1}{(4 \pi D t)^{1/2}}
\cos \left(\omega_L t
 \left[ 1 + \frac{ 2 \tau_0 \ln|\omega_L \tau_0|}{\pi \tau^*} \right] \right)
\exp\left[-\frac{\omega_L \tau_0 t}{\tau^*} - \frac{t}{\tau_s} - \frac{L^2}{4 D t} \right]
.
\end{equation}
\end{widetext}
Comparing Eq. (\ref{resulting}) to Eq. (\ref{ClassicHanle}), we find that they have the {\em same analytical structure} and can be reduced to each other upon replacement
\begin{align}
\label{effective}
\frac{1}{\tau_s} &\rightarrow  \frac{1}{\tau_s} + \frac{\omega_L \tau_0}{\tau^*} = \frac{1}{\tilde \tau_s},\\
{\omega}_L &\rightarrow \frac{2}{\pi} \omega_L \left( \frac{\tau_0}{\tau^*} \right)
\ln\frac{1}{ \omega_L \tau_0} = {\tilde \omega}_L.
\label{effective1}
\end{align}
While the integral Eq. (\ref{resulting}) can be evaluated analytically for arbitrary distance $L$ between the electrodes, the effect of traps on the shape of the Hanle profile is most pronounced in the limit of short channel $L\ll (D\tau_s)^{1/2}$. In this limit, the bare
shape Eq. (\ref{ClassicHanle}) simplifies to
\begin{equation}
\label{asymptotic}
R_H(\omega_L) \propto \frac{\sqrt{\sqrt{1+ { \omega}_L^2 { \tau}_s^2} + 1}}{\sqrt{1+ { \omega}_L^2 { \tau}_s^2}},
\end{equation}
and depends only on the product $\omega_L\tau_s$.
In the presence of traps, this product should be
replaced  by
\begin{equation}
\label{nontrivial}
{\tilde \omega}_L {\tilde \tau}_s = \frac{2}{\pi} \ln\left[ \frac{1}{\omega_L \tau_0} \right] \frac{\omega_L \tau_s }{\omega_L \tau_s + \frac{\tau^*}{\tau_0}}.
\end{equation}
Two important messages can be inferred from Eq. (\ref{nontrivial}):
({\em i}$\,$)~the presence of traps leads to the
{\em narrowing} of the Hanle curve from $\omega_L \sim \frac{1}{\tau_s}$ to
$\omega_L \sim \frac{1}{\tau_s}\left(\frac{\tau^*}{\tau_0}\right)$,
({\em ii}$\,$)~for higher fields the Hanle curve are completely {\em flat}.
%It follows from Eq. (\ref{nontrivial}) that the presence of traps leads to the
%narrowing of the Hanle curve from $\omega_L \sim \frac{1}{\tau_s}$ to
%$\omega_L \sim \frac{1}{\tau_s}\left(\frac{\tau^*}{\tau_0}\right)$.
The suppression of the widths is given by the
ratio, $\tau^*/\tau_0$, of the diffusion time
between the traps to the trapping time. The dependence of this factor on
the density of traps is ${\cal N}^{-2}$, as it follows from Eq. (\ref{InOtherWords}). Narrowing of the Hanle curves with ${\cal N}$ is illustrated in Fig. \ref{fig3}.
Note that in Eqs. (\ref{effective1}), (\ref{nontrivial}) we have already 
set $\tau^*$ to be much less than $\tau_0$, so that our result Eq. (\ref{nontrivial}) already assumes that the narrowing of the Hanle curve is substantial. 

%In deriving Eq. (\ref{nontrivial}) we assumed that $t\ll t_{\s tr}$.
%Since the time $t_{\s tr}$ can be estimated as $t_{\s tr} \sim n(t)\tau_0\approx t\tau_0/\tau^{\ast}$, the above condition reduces to $\tau_0\gg \tau^{\ast}$, i.e. the time spent on a typical trap is longer than the inter-trap traveling time. It is under this condition that the substantial  narrowing of the Hanle profile takes place.

%This condition can be expressed as
%have  adopted two assumptions. Firstly, we assumed that the carrier spends most of the time sitting on the trap, and thus neglected
%$t$ in the argument of Eq. (\ref{withdeltafunction}). We can now express this condition quantitatively, as
%$\Tau_0\Gg \Tau^{\Ast}$, I.E. The Time Spent On A Typical Trap Is Longer
%Than The Inter-Trap Traveling Time.
%On The Other Hand,
%The time $t_{\s tr}$ can be estimated as
%second term in the argument of cosine in Eq. (\ref{withdeltafunction}) is of the order of
%$n(t)\tau_0\approx t\tau_0/\tau^{\ast}$, so it is indeed bigger than $t$.

We have also assumed that the characteristic width of the Hanle profile
%Eq. (\ref{smallL})
is much smaller than $\tau_0^{-1}$. With the width given by  $\omega_L \sim \frac{1}{\tau_s}\left(\frac{\tau^*}{\tau_0}\right)$, the above condition reduces to
$\tau^*\ll \tau_s$,
i.e. the loss of the spin memory on the way between two neighboring traps is small. This condition is implicit for our scenario, since we presumed that the number
of visited traps is big.

%\section{Discussion}
\noindent{\em Discussion}.
The authors of Refs. \onlinecite{NoInjection1,NoInjection2,NoInjection3} arrived at the conclusion that the Hanle effect in organic spin valves is missing
%The conclusion that the Hanle effect in organic spin valves is missing
%%\cite{NoInjection1,NoInjection2,NoInjection3}
%was drawn in Refs. \onlinecite{NoInjection1,NoInjection2,NoInjection3}
on the basis of the following measurements. The difference, $\Delta {\cal R}$, between
the resistances for parallel and antiparallel orientation of magnetizations of electrodes was measured under the conditions when one of the electrodes, CoFe\cite{NoInjection2} or Co\cite{NoInjection3}, was close to the magnetization reversal. For the orientation of the external magnetic field, ${\bm \omega}_L$, normal to both magnetizations, measured $\Delta {\cal R}$ did not depend on ${\bm \omega}_L$. If the conventional Hanle effect was at work, the value $\Delta {\cal R}$ would vanish with ${\bm \omega}_L$. This is because, the stronger is ${\bm \omega}_L$,
the weaker is the memory of the carrier arriving at the fully magnetized LSMO\cite{NoInjection1,NoInjection2,NoInjection3} electrode about its initial
spin direction.

Theoretically, the decay of $\Delta{\cal R}$ with ${\bm \omega}_L$ is described by
Eq. (\ref{asymptotic}) and is shown in Fig. \ref{fig3} with a solid line. In Fig. \ref{fig3} (dashed lines) we also
see that in the presence of traps the Hanle curve does not drop,
but stays  {\em flat} except for a narrow domain of small fields. This plateau behavior would account for the observations  of Refs. \onlinecite{NoInjection1,NoInjection2,NoInjection3}. Concerning a narrow peak,
if its width is smaller than the resolution in ${\bm \omega}_L$,
it would not show up. This resolution can be set e.g. by the earth's magnetic field $\sim 0.1$mT.

Here we emphasize that the overall shape of the Hanle curve, in the presence of traps, which is a  narrow peak on top of a plateau, is a direct consequence of the broad waiting-time distribution. Without a spread in the waiting times, the traps would simply lead to
a homogeneous  narrowing of a standard Hanle profile Eq. (\ref{asymptotic}) by a factor $\tau^*/\tau_0$. This, in turn, would mean that $\Delta {\cal R}$ drops to
zero for the applied fields $\omega_L \gtrsim \frac{1}{\tau_s}\left(\frac{\tau^*}{\tau_0}\right)$. Thus the unique independence of
$\Delta {\cal R}$ of $\omega_L$, which is in line with experimental findings, can be traced to the heavy-tailed distribution of the trapping times.

%\section{Concluding Remarks}
\noindent{\em Concluding Remarks}:
%\begin{itemize}

%\item
\noindent\textbullet~
Our main finding  that, with spin-preserving traps, the GMR and the Hanle effects become ``decoupled" from each other can be elaborated on as follows.
The relation $\lambda_s =(D\tau_s)^{1/2}$  no longer holds in the presence of traps.
 The value $\lambda_s$  determined from the thickness dependence of GMR, as in Refs. \onlinecite{OrganicValve1'},  \onlinecite{ValyIsotopes}, does not ``know" about the traps. At the same time, the effective ${\tilde\tau}_s$ defined by Eq. (\ref{effective}), which governs the Hanle profile, does.

\noindent\textbullet~
In the toy model we assumed that the microscopic mechanism of the spin memory loss are the on-site random field. In fact, the origin  of $\tau_s$ in  Eq. (\ref{asymptotic}) for the Hanle profile can be both, random hyperfine fields and spin-orbit interactions.

%In short, the degree to which the trapping affects the charge transport can be quantified via parameter $\tau^*/(\tau^*+\tau_0)$, which is the portion of time during which the carrier actually travels rather than sits on the traps.   Our calculation shows that, when this ratio is small, the Hanle curve narrows by this ratio.
%
% {\bf Relation $\lambda_s =(D\tau_s)^{1/2}$ does not hold in the presence of traps.
%If $\lambda_s$ is determined from the thickness dependence of GMR, it does not know about the traps, while the effective $\tau_s$ in Hanle does.}
%

%\item
\noindent\textbullet~
The strong assumption which underlies the decoupling of the GMR and the Hanle effects,
adopted in the present manuscript, is that spin-memory is not lost while
the carrier sits on the deep trap.
This, in turn, requires that the wave function of the trap state does not overlap
with hydrogen protons.
In experiments Refs. \onlinecite{NoInjection1,NoInjection2,NoInjection3}, the organic layers of spin valves were based on Alq$_3$ and PTCDI-C4F7 organic molecules. The  hydrogen atoms in Alq$_3$ are attached to approximately $50\%$ of carbon atoms and their locations are well studied\cite{alq3}. It is also accepted that traps play a
prominent role in transport through organic layers\cite{traps,traps1}.
However, the spatial positions of the fragments of Alq$_3$ molecules repsonsible
for the trap states are not known.
%the spatial positions of the trap states in Alq$_3$ were not reported.
%This can be the case due to
%the smallness of the size of the wave function of the trap state.
%Random hyperfine field leading to the spin dephasing   is proportional
%to the square root of the number of protons ``in contact" with the carrier spin.
Note also, that in our consideration  we have completely  neglected the effect of pairs of traps.\cite{FlatteTrap}

%\item
%The regime of ``long" device $L^2\gg D\tau_s$ seems adequate for
%organic devices due to large $\tau_s$, which, from independent measurements (Baker) comes out $10^3$ times longer than in conventional semiconductors,
%while diffusion coefficient, although not known accurately, can be estimated from the transit time {\bf Schmidt} and  is many orders of magnitude smaller than in silicon.
%{\bf Optical
% $\tau_s$ can be much longer than $\tau_s$ in Hanle}

%\item
\noindent\textbullet~
Obviously, the flat shape of the Hanle curve in Fig. \ref{fig3} applies only
in a finite field domain. Indeed, in deriving Eqs. (\ref{asymptotic}),
(\ref{nontrivial})
we treated the product $\omega_L\tau_0$, which is the precession angle on a single trap, as a small parameter. It is intuitively clear that for $\omega_L\tau_0 \gg 1$ the Hanle curve should decay. What is surprising, is that this decay is very slow. To capture it analytically, one has to take the $\omega_L\tau_0 \gg 1$ limit of Eq. (\ref{wide}) and use it in Eq. (\ref{resulting}) instead of the low-field expansion.
The result amounts to replacement $\omega_L\tau_0$ by $\ln(\omega_L\tau_0)$ in the exponent, and also to the replacement of the argument of cosine by $t/\tau^*$. This, in turn, leads to the following modification of the Hanle curve
Eq.~(\ref{asymptotic}): the product $\omega_L\tau_s$ gets replaced by $\ln(\omega_L\tau_0)$. We see that $R_H(\omega_L)$ {\em does} decay at strong enough fields, but this decay is logarithmical, i.e. very slow.

%\item
%An extravagant explanation of the absence of the Hanle effect in organic spin-transport devices
%was offered by Z. G. Yu in Ref. \onlinecite{Yu}, and according to which the Hanle curve is simply too broad to be
%observable. To substantiate his claim, Yu conjectures that, in polymers, the exchange interaction of two electrons of neighboring hopping sites is huge. As a result, the diffusion coefficient in
%the spin transport equation is much bigger that the diffusion coefficient of the current carrier.
%Based on that, Yu argues that it takes a real strong magnetic field to induce the spin precession. Yu's message is similar to voltage buildup in magnetic
%insulator due to the flow of spin-waves {\bf references} . This buildup requires conversion of
%spin current into the charge current, i.e. inverse spin Hall effect.

\noindent\textbullet~
The essence of the explanation\cite{Yu} of missing Hanle effect in organic structures
is assumption that the diffusion coefficient in
the spin-transport equation is much bigger than the diffusion coefficient of a current carrier. This assumption is attributed to a strong exchange interaction of two carriers of neighboring hopping sites, so that the spin polarization is sensed by the detector  much faster than the injected charge actually reaches it. This makes the spin transport robust to the external field.
Such a ``spin-wave" scenario is similar to the voltage buildup in magnetic
insulator due to the flow of spin-waves\cite{SpinSeebeck1,SpinSeebeck2}, and seems questionable
since the voltage buildup requires conversion of
spin current into the charge current, i.e. inverse spin Hall effect.

%seems questionable to us, since the Hanle effect is detected by voltage
%buildup, which is the result of carrier diffusion rather that pure spin diffusion.
%In addition, the values $\tau_s$ inferred from non-transport measurement do not
%suggest that characteristic magnetic field $\omega_L\sim 1/\tau_s$ for Hanle is anomalously high.

%\end{itemize}

%\section{Acknowledgements}
\noindent{\em Acknowledgements}.
We are grateful to Z. V. Vardeny for reading the manuscript and  providing illuminating remarks. We acknowledge interesting discussions of spin transport with V. V. Mkhitaryan.
We have  also benefited from discussions with 
C. Boehme, J. M. Lupton, and A. Tiwari concerning different aspects of the Hanle effect. This work was supported by NSF
through MRSEC DMR-1121252.

\end{document}